\documentclass[a4paper]{spie}
\usepackage[utf8]{inputenc} 
\usepackage[T1]{fontenc}
\usepackage{graphicx,xspace,amsmath,amssymb,float}
\usepackage{ifpdf,epstopdf}
\usepackage{xparse}
\usepackage{textcomp}
\usepackage{lastpage}

\usepackage[dvipsnames]{xcolor}
\definecolor{mydarkblue}{rgb}{0., 0., 0.35}
\usepackage[pdfa]{hyperref}
\hypersetup{colorlinks=true,allcolors=mydarkblue} 

\usepackage{xparse} 
\usepackage[normalem]{ulem} 
\usepackage{soulutf8} 

\usepackage{fancyhdr}
\renewcommand{\headheight}{25pt} 
\renewcommand{\headsep}{40pt} 

\newcommand{\bfepsilon}{\boldsymbol{\epsilon}}
\newcommand{\bfh}{\boldsymbol{h}}
\newcommand{\bfhtilde}{\tilde{\bfh}}
\newcommand{\bfi}{\boldsymbol{i}}
\newcommand{\bfo}{\boldsymbol{o}}
\newcommand{\bfn}{\boldsymbol{n}}
\newcommand{\bfS}{\boldsymbol{S}}
\newcommand{\bfx}{\boldsymbol{x}}
\newcommand{\bfy}{\boldsymbol{y}}
\newcommand{\Bb}{\boldsymbol{B}}

\newcommand{\ddroit}{\mathrm{d}}
\newcommand{\diag}[1]{\mathrm{diag}\left(#1\right)}
\newcommand{\eg}{\emph{e.g.},\xspace}

\newcommand{\ie}{\emph{i.e.},\xspace}
\newcommand{\matC}{\boldsymbol{C}}
\newcommand{\matF}{\boldsymbol{F}}
\newcommand{\matH}{\boldsymbol{H}}
\newcommand{\matM}{\boldsymbol{M}}
\newcommand{\matS}{\boldsymbol{S}}
\newcommand{\matW}{\boldsymbol{W}}

\newcommand{\PIC}{\textsc{pic}\xspace}

\newcommand{\SPIDER}{\textsc{spider}\xspace}

\newcommand{\tfo}{\tilde{\bfo}}
\newcommand*{\QED}[1][$\square$]{%
\leavevmode\unskip\penalty9999 \hbox{}\nobreak\hfill
    \quad\hbox{#1}}

\graphicspath{{./Figures/}}

\begin{document}

\title{Analysis of a compact interferometric imager}
\author[a]{L.\ M.\ Mugnier}
\author[b]{V.\ Michau}
\author[a]{H.\ Debary}
\author[c]{F.\ Cassaing}
\affil[a]{{DOTA, ONERA, Université Paris Saclay\\
29 Av.\ de la Division Leclerc, F-92322 Châtillon, France}}
\affil[b]{{DSG, ONERA, Université Paris Saclay\\
         6 Ch.\ de la Vauve aux Granges, F-91120 Palaiseau, France}}
\affil[c]{{DTIS, ONERA, Université Paris Saclay\\
         6 Ch.\ de la Vauve aux Granges, F-91120 Palaiseau, France}}

\authorinfo{Further author information: \\E-mail: mugnier@onera.fr}


\maketitle

\begin{abstract}
The advent of photonic integrated circuits (PICs) will allow the replacement of the large aperture of an optical telescope by a dense array of small apertures combined interferometically. The light coming from aperture pairs can be combined by a PIC in order to extract interferogram characteristics known as complex visibilities, from which the observed object can then be reconstructed. In such a compact interferometric imager, the optical components dedicated to image formation in a regular telescope are no longer necessary. In particular, such a concept is relevant for space missions where weight and size are critical.

In this communication, we study such an instrument concept, focusing on signal-to-noise considerations. We recall the design basis for the field and the spatial resolution, and we show that the spectral resolution must be no less than the field to resolution ratio. Then, we analyze the signal-to-noise ratio of this concept, assuming that each spatial frequency is recorded only once, and compare the signal-to-noise ratio with that of a monolithic telescope. We perform the comparison in Fourier space for an identical number of recorded photons. We show that the noise propagation of the interferometric imager is identical to that of a monolithic telescope that would have a flat Modulation Transfer Function 
with a level roughly given by the ratio of the small apertures’ diameter to the maximum baseline. We conclude that the noise propagation in low and medium spatial frequencies is unfavorable for the interferometric imager.
\end{abstract} 


\section{Description of the SPIDER concept}
\label{chap-concept-spider}

\begin{figure}[htbp]
\begin{center}\leavevmode
  \includegraphics[width=.75\linewidth]{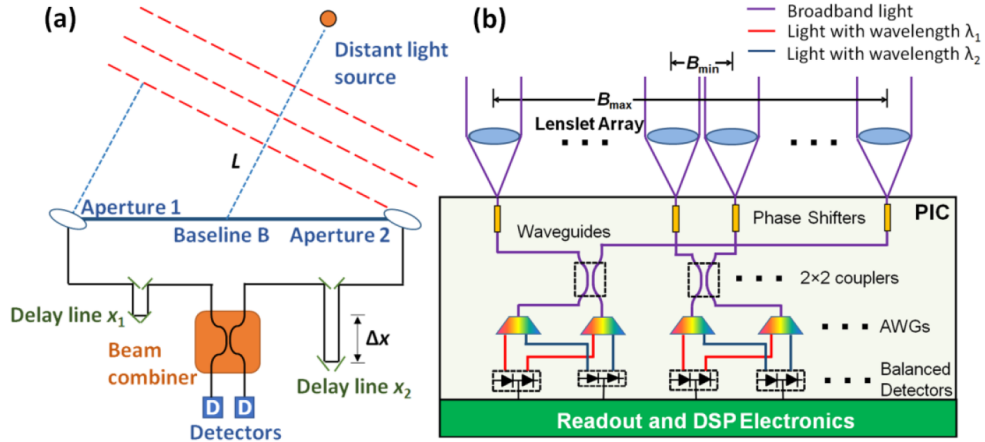}
  \caption{a: principle of an interferometer (long baseline, delay lines, fibered combination); b: principle of \SPIDER (micro-lens array, injection in a PIC that contains phase shifters, couplers, dispersing elements (AWG) and detectors). Illustration taken from Ref.~\citeonline{su_interferometric_2018}.
  \label{fig-interferometre-vs-spider}
  }
  \end{center}
\end{figure}

The innovative optical imaging system concept based on interferometry known as \SPIDER\footnote{\SPIDER stands for Segmented Planar Imaging Detector for Electrooptical Reconnaissance.}~\cite{kendrick_segmented_2013} could bring substantial gains in size and weight compared to a conventional focal plane imager. 
%
This compact interferometric imager concept combines the following ideas:
\begin{itemize}
    \item replace a focal plane imager by an interferometer one may view as a reduced model of an astronomical interferometer, which is illustrated Fig.~\ref{fig-interferometre-vs-spider}a~;
\item avoid the long-stroke delay lines of Fig.~\ref{fig-interferometre-vs-spider}a by having the set of apertures on a common mount;
\item use the technology of \PIC{}s (\emph{Photonic Integrated Circuit}) to realize in an extremely reduced thickness the functions of phase shifting, beam coupling, spectral dispersion and detection, as illustrated Fig.~\ref{fig-interferometre-vs-spider}b.
\end{itemize}
The concept of \SPIDER was presented in 2013~\cite{kendrick_segmented_2013}, by researchers at Lockheed-Martin, and several preliminary experimental demonstrations followed~\cite{scott_demonstration_2014,su_experimental_2017}.
Several aspects of such a concept have been studied since then~\cite{liu_system_2018,gao_quantitative_2018,%
guo-mian_improved_2019,cao_lenslets_2020,liu_system_2020}.
%
A $Si_3N_4$ PIC demonstrator with 12 baselines and 18 spectral channels dispersed by an \emph{Arrayed Waveguide Grating} or AWG was then developed by Lockheed-Martin and UC Davis~\cite{duncan_spider_2016}. The 
experimental laboratory demonstration of the near-infrared SPIDER imager based on this PIC resulted in a publication in 2018~\cite{su_interferometric_2018}, which is the most comprehensive publication by these two teams.

The diagram in figure~\ref{fig-interferometre-vs-spider} shows that at their entry in the PIC, the beams in each guide meet first the phase shifters, which must therefore be achromatic, then the couplers, which combine the apertures two by two, then only the dispersive elements (AWG), and finally the detectors.
The flux recorded by the detectors allows, via a modulation produced by the phase shifters, to reconstruct the contrast and the phase of the interference fringes of each pair of recombined apertures. The contrast and phase are grouped into a  quantity called \emph{complex visibility},
which gives a sample of the Fourier Transform of the object via the 
Van Cittert-Zernike theorem. Then, the object can be estimated through an image reconstruction, using a method among those developed in astronomical interferometry~\cite{Besnerais-a-08,Thiebaut-a-17}. 

\begin{figure}[H] 
\begin{center}\leavevmode
  \includegraphics[width=.75\linewidth]{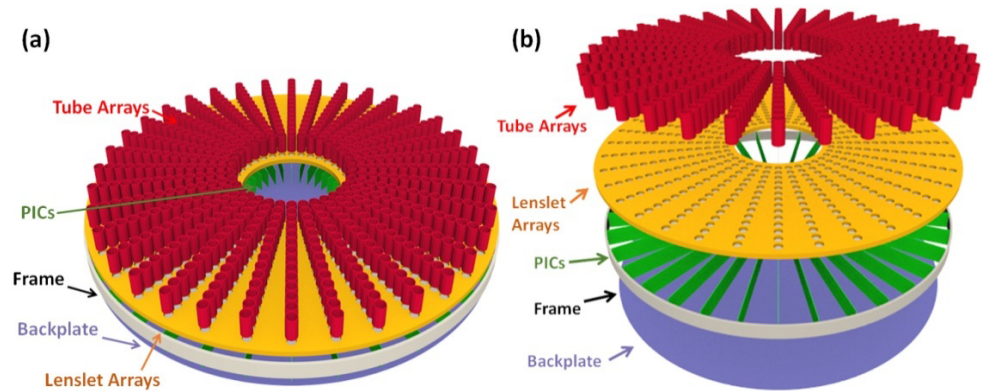}
  \caption{schematic diagram of a compact interferometric imager (taken from Ref.~\citeonline{su_interferometric_2018}).
  \label{fig-schematic-spider}
  }
  \end{center}
\end{figure}

The telescope uses a large number of baselines to sample the complex visibility of the observed object on a rather large number of spatial frequencies.  
Figure~\ref{fig-schematic-spider}, taken from Ref.~\citeonline{su_interferometric_2018}, shows the schematic diagram of such a telescope. 

\clearpage
\section{Physical analysis and basic design rules }\label{chap-dimensionnement}

\subsection{Field of view}\label{sec-champ}
The field-of-view (FOV) of an interferometric imager such as \SPIDER, as for any pupil-plane interferometer, is limited to $\lambda/D$, where $\lambda$ is the central wavelength considered and $D$ is the diameter of the apertures. Moreover, because of the injection in waveguides, the observed object is apodized by the ``antenna lobe'' in intensity $L$ of the waveguide into which the light is injected, so that the effective object whose complex visibilities are measured by the interferometer is 
\begin{equation}
    o_{\text{eff}}(\alpha_x,\alpha_y)= o(\alpha_x,\alpha_y) \times L(\alpha_x,\alpha_y) .\label{eq-o-effectif}
\end{equation}
Finally, we note that  $L$ is the intensity of the back-propagation towards the object of a pupil amplitude $P$, which is the waveguide's propagation mode in the aperture plane: $L= |\mathrm{FT}(P)|^2$. $L$ is typically a Gaussian 2D function of FWHM $\lambda/D$, as studied in details in Refs.~\citeonline{dyer_pupil-size_1999,mege_interferometry_2003}. The FOV $\theta_{\text{fov}}$ seen by the instrument is thus:
\begin{equation}
   \theta_{\text{fov}} \simeq \lambda/D 
   \;.
\end{equation}

\subsection{Resolution and number of resolution elements}
As for any interferometer, the maximum spatial frequency recorded by a compact interferometric imager is given by the maximum distance separating two apertures, or maximum baseline $B_{\text{max}}$, and is $B_{\text{max}}/\lambda$. 
The angular resolution of such an instrument is thus
\begin{equation}
   r = \lambda/B_{\text{max}} \;. 
\end{equation}
The angular field being $\lambda/D$, the number of resolved elements of a compact interferometric imager in the observable angular field is $B_{\text{max}}/D$.
For an object sampling fulfilling the Shannon-Nyquist criterion, this means that the reconstruction will give an object on a support of side $N_{\text{Shannon}}$ pixels given by:
\begin{equation}
   N_{\text{Shannon}} = 2 B_{\text{max}}/D  \;. 
\end{equation}

\subsection{Spectral width}
The spatial frequency of the object measured for point-like apertures separated by $\Bb$ interfering at a wavelength $\lambda$ with zero bandwidth is $\Bb/\lambda$. As soon as the bandwidth is non-zero, we record in the interferogram an average of spatial frequencies of the object for all the wavelengths of the band. Therefore, what spectral bandwidth can be tolerated?

To answer this question, it is useful to realize that Equation~\ref{eq-o-effectif} means that even at zero spectral width, an interferometer with waveguide injection intrinsically measures an average of spatial frequencies of the object over a typical width of $D/\lambda$. Another way of picturing this is to see that the recombined beams contain interferences between apertures of diameter $D$ spaced by $B$ and thus between pairs of points separated by distances ranging from $(B-D)$ to $(B+D)$.

It is then reasonable to specify that the spatial frequency averaging due to the spectral width $\ddroit\lambda$ is less than the intrinsic averaging due to the non-zero aperture size. Differentiating $f=B/\lambda$ we obtain the following condition:%
\footnote{Interestingly, this condition can also be obtained by examining the temporal coherence, and specifying that the opd between two apertures separated by $B$ for a direction $\theta=\lambda/(2D)$ at the edge of the field of view, which is $\text{opd}=B.\theta=B\lambda/(2D)$, must be less than half the coherence length $L_c=\lambda^2/(\ddroit\lambda)$. }
\begin{equation}
    \ddroit f = \frac{B}{\lambda^2}\ddroit\lambda < D/\lambda.\nonumber
\end{equation}
For the shortest wavelength and the longest baseline, this yields:
\begin{equation} \label{eq-resol-spectrale}
   \frac{\ddroit\lambda}{\lambda_{\text{min}}} < \frac{D}{B_{\text{max}}} \;.
\end{equation}
Inverting this inequality, we can also read it as follows: the spectral resolution ${{\lambda}/{\ddroit\lambda}} $ must be greater than the number of (spatially) resolved points of the interferometer, $B_{\text{max}}/D$.


\section{Physical modeling for performance estimation}
\label{chap-modele-physique}

\subsection{Data model}

Let us consider a pair of apertures separated by $\Bb$ and thus measuring the spatial frequency $(u,v)=\Bb/\lambda$, and let us call $C_1$ (respectively $C_2$) the coupling coefficient of aperture 1 (respectively 2) towards the output. Let $\phi_k$ be the modulation phase of the $k$-th measurement, and $N_{\text{ph}}$ the mean number of photons received in each 
of the two apertures; then, assuming balanced channels \ie $C_1=C_2=1/\sqrt{2}$,
a model of the recorded raw data (intensities) is

\begin{equation}
    i(u,v,\phi_k) = N_{\text{ph}} \left\{1 + |\gamma(u,v)| \cos[\phi_k + \theta(u,v)] \right\} ,\label{eq-mesure-i}
\end{equation}
where the complex visibility of the observed object is:
\begin{equation}
\gamma(u,v) = |\gamma(u,v)| \exp(j\theta(u,v)) \;.
\end{equation}


By using typically four measurements
($\phi_k=(k-1)\pi/2$, for $k\in\{1,2,3,4\}$) followed by a demodulation\footnote{For each spatial frequency $(u,v)$, the demodulation consists in estimating $y(u,v)$ as: $\Re(y(u,v)) = (i(u,v,\phi_1) - i(u,v,\phi_3))/2$ and $\Im(y(u,v))=(i(u,v,\phi_4) - i(u,v,\phi_2))/2$.}, one 
obtains a complex measurement $y(u,v)$ proportional to the mean number of photons received in each of the two apertures and to the complex visibility of the object at the measured spatial frequency:
\begin{equation}
 y(u,v) = N_{\text{ph}} \,|\gamma(u,v)| \exp(j\theta(u,v)) = N_{\text{ph}} \gamma(u,v) \;.\label{eq-y-complexe}
\end{equation}
Although it does not change the global photon budget, here we assume that (only) two raw measurements are obtained simultaneously using the two outputs in phase opposition of each coupler. Two other measurements must then be obtained by introducing a phase shift of $\pi/2$. Each complex measurement thus uses $4N_{\text{ph}}$ photons (2 raw interferometric measurements, each receiving $N_{\text{ph}}$ photons per aperture).

Finally, let us linger for a moment on the flux of this $\bfo$ object, which can be reconstructed from the complex measurements $y(u,v)$ above. We have: \begin{equation}
\gamma(u,v) = \tfo(u,v) / \tfo(0,0)
\end{equation}
thanks to the Van Cittert-Zernike theorem, so $\gamma(0,0)=1$.
%
Equation~\ref{eq-y-complexe} then yields $y(0,0)=N_{\text{ph}}$, \ie the  collection $\bfy$ of complex measurements $y(u,v)$ is a set of samples of the Fourier transform of an object $\bfo$, of integral (\ie of total flux) $N_{\text{ph}}$: 
\begin{equation}
\int o(x,y)\,\ddroit{}x\ddroit{}y=\tfo(0,0)=N_{\text{ph}}\;. \label{eq-integrale-o}
\end{equation}

\subsection{Noise modeling and comparison with a conventional imager}
%
A complete physical modeling of the compact interferometric imager concept must include the modeling of noise, and if possible comparatively to a classical \ie focal plane imager. The propagation of noise in the algorithm that reconstructs the object from complex visibilities is delicate because
these algorithms can be notably non-linear. In order to avoid dealing with such noise propagation, we have chosen to model the noise of a compact interferometric imager and of a classical instrument in the Fourier \ie spatial frequency domain.

We assume in the following that the measurement noise is predominantly photon noise, which is reasonable for many scenarios at least in the visible and near-infrared. In the noise analysis, we additionally assume that the compact interferometric imager measures each spatial frequency only once, as in Ref.~\citeonline{Debary-p-22}.

\subsubsection{Noise modeling for a compact interferometric imager}

%

For a compact interferometric imager, the variance of the 
noise on the complex measurement $y(u,v)$ of Equation~\ref{eq-y-complexe}, resulting from the demodulation of four raw measurements given by  Equation~\ref{eq-mesure-i}, is\footnote{The variance of the complex variable $y$ can be defined as $\mathrm{Var}(y)\triangleq E(|y - E(y)|^2)$, and it is easy to show that $\mathrm{Var}(y) = E(|y|^2) - |E(y)|^2 
= \mathrm{Var}(\Re(y)) + \mathrm{Var}(\Im(y))$. If moreover, as we can reasonably assume in this study, $\Re(y)$ and $\Im(y)$ are two independent Gaussian variables of the same variance, then $y$ is a complex circular Gaussian variable (distribution invariant by any rotation).}:
\begin{equation}
    \sigma_y^2(u,v)=  N_{\text{ph}} , 
\end{equation}
where $N_{\text{ph}}$ is still the average number of photons received in each of the two apertures contributing to the interference during a raw measurement.
This directly follows from the fact that $y(u,v)$ is the demodulation of 4 raw data measurements, where each is corrupted by photon noise, of variance $N_{\text{ph}}$ under the assumption that $|\gamma(u,v)| \ll 1$.

The standard deviation of these complex measurements, normalized by the object's Fourier transform at the zero frequency, is therefore simply, according to Equation~\ref{eq-integrale-o}:
\begin{equation}
    \boxed{\frac{\sigma_y(u,v)}{\tfo(0,0)} = \frac{1}{\sqrt{N_{\text{ph}}}} } \;.
    \label{eq-erreur-normalisee-Spider}
\end{equation}
%
The set of spatial frequencies to be measured to cover the same frequency support as a conventional imaging instrument of the same resolution, \ie of diameter $B_{\text{max}}$, is contained in a half-disk of 
radius
$B_{\text{max}}/\lambda$, that is to say of the order of $\pi(B_{\text{max}}/D)^2/2$ distinct frequencies, if the frequency sampling step is $D/\lambda$ (which requires joint apertures).

The measurement of all these spatial frequencies therefore takes a total of
$\pi(B_{\text{max}}/D)^2/2 \times 4 N_{\text{ph}}$ photons:
\begin{equation}
    N_{\text{phtot}} = 2\pi(B_{\text{max}}/D)^2 N_{\text{ph}} .\label{eq-Nphtot}
\end{equation}


\subsubsection{Noise modeling for a conventional imager}

%
For a classical imaging instrument, we additionally assume that the observed scene is of sufficiently homogeneous luminance, so that the noise can be assumed to be stationary, \ie so that the variance of the noise of the classical imager $\sigma_n^2$ can be assumed to be approximately constant over all $N_{\text{pix}}$ pixels.
%
The discretized object that yields the recorded digital image $\bfi'$ is noted here as $\bfo'$ to distinguish it from the object in the compact interferometric imager model, since the total flux of $\bfo'$ is different:
\begin{equation}
    \sum_{p,q} o'(p,q) = N_{\text{phtot}} ,
\end{equation}
%
where $N_{\text{phtot}}$ is the total number of photons received (on average) during the recording of the image $\bfi'$.
%
The image model writes:
\begin{equation}
    \bfi' = \bfh \star \bfo' + \bfn ,
\end{equation}
where $\bfh$ is the discrete PSF, and $\star$ is the discrete convolution operator.

We restore this object, whose prior law is taken second order stationary and with Energy Spectral Density (ESD) $\bfS_o$, by a Wiener filter. We then examine the propagation of the image noise on the discrete Fourier transform (DFT) of the restored object $\hat{\bfo}$, where the DFT of an image $\bfx$ is defined as:
\begin{equation}
    \left[\mathrm{DFT}(\bfx)\right](p',q') = \sum_{p,q} x(p,q) \, \exp(-2j\pi[p.p'+q.q']/N_{\text{pix}}) ,\label{TFD}
\end{equation}
\emph{without a factor $1/N_{\text{pix}}^2$} (as in some languages, \eg IDL), and noted $\matF\bfx$ in matrix form.
Let $\bfepsilon=\hat{\bfo}'-\bfo'$ be the restoration error of the object, we show in appendix~\ref{annexe-matcoverr} that the covariance matrix of the DFT of this error, noted $\Gamma_{\matF\bfepsilon}$, is diagonal and given by:
\begin{equation}
    \boxed{\Gamma_{\matF\bfepsilon} = 
    \diag{\frac{N_{\text{pix}}^2 \sigma_n^2}%
         {|\bfhtilde|^2 + \frac{N_{\text{pix}}^2 \sigma_n^2}{\bfS_o}}} } \;.
         \label{eq-mat-cov-err-F}
\end{equation}
The interpretation of 
Equation~(\ref{eq-mat-cov-err-F}) is easy: 
the numerator represents the variance of the noise in the image DFT at any frequency so in particular at zero frequency, and this variance is therefore the sum of the variances of the noise on all pixels, i.e. $N_{\text{pix}}^2 \sigma_n^2$. 
The denominator represents the amplification of the noise by the restoration and its form is reminiscent of the Wiener filter: FTM squared plus the ratio between the variance of the noise in the image DFT and the object's ESD. Finally, the fact that the $\Gamma_{\matF\bfepsilon}$ matrix is diagonal means that the restoration error is decorrelated in the Fourier domain between spatial frequencies, which comes jointly from the stationary character of the noise and the stationary prior probability distribution of the object.

Since we have assumed the photon noise to be dominant, this can further be written as:
\begin{equation}
    \Gamma_{\matF\bfepsilon} = 
    \diag{\frac{N_{\text{phtot}}}%
         {|\bfhtilde|^2 + \frac{N_{\text{phtot}}}{\bfS_o}}} \;.
\end{equation}
In particular, at spatial frequencies for which noise is not dominant, we see that the variance of the restoration error in the Fourier domain is simply equal to the variance of the noise in the DFT of the image divided by the MTF squared. In other words, the noise in the Fourier domain is, in standard deviation, simply amplified by the inverse of the MTF.
By normalizing this standard deviation by the DFT of the object at zero frequency, we obtain, at spatial frequencies for which signal dominates noise:
\begin{equation}
    \boxed{\frac{\sigma_{\matF\bfepsilon}}{\tfo'(0,0)} \simeq \frac{1}{|\bfhtilde|\sqrt{N_{\text{phtot}}}} }\;.
    \label{eq-erreur-normalisee-imageur}
\end{equation}

\subsubsection{Comparison}\label{sec-comparRSB}

To compare the noise propagation between a classical imager and a compact interferometric imager, we will assume that the classical imager receives $N_{\text{phtot}}$ photons in a single image, where $N_{\text{phtot}}$ is given by Equation~(\ref{eq-Nphtot}), so that
%
Equation~(\ref{eq-erreur-normalisee-imageur}) can be re-written this way:
\begin{equation}
    {\frac{\sigma_{\matF\bfepsilon}}{\tfo'(0,0)} 
    = \frac{1}{\sqrt{2\pi} \,|\bfhtilde|\,(B_{\text{max}}/D) \sqrt{N_{\text{ph}}} } }\;.
    \label{eq-erreur-normalisee-imageur2}
\end{equation}
Comparison of Equations~\ref{eq-erreur-normalisee-Spider} and~\ref{eq-erreur-normalisee-imageur2} suggests that the noise propagation is similar for both types of instruments, \ie proportional to $1/\sqrt{N_{\text{ph}}}$, and that it is identical if:
\begin{equation}
    |\bfhtilde| = \frac{1}{\sqrt{2\pi}} \frac{D}{B_{\text{max}}} ,
    \label{eq-comparaison-bruit}
\end{equation}
at all frequencies except the zero frequency, at which the MTF is 1 by convention. In other words, with respect to noise propagation, a ``standard''
compact interferometric imager (standard in the sense that each spatial frequency would be measured once and only once) is equivalent to a conventional imager with a flat MTF of the order of ${D}/{B_{\text{max}}}$, \ie the inverse of the number of apertures in the maximum baseline of the instrument

This result can also be rephrased as follows: the compact interferometric imager studied here is equivalent, with respect to noise propagation, to a focal plane multi-aperture imaging instrument with an aperture that is a phased array of non-redundant apertures, and comprising approximately
\begin{equation}
    N_{\text{tel}}=\sqrt{2\pi}B_{\text{max}}/D \label{eq-Ntel}
\end{equation}
apertures.
Indeed we know that, for an ideal non-redundant phased array instrument, the MTF is flat and equals $1/N_{\text{tel}}$, where $N_{\text{tel}}$ is the number of sub-apertures. 

Finally, the above results show that, for the highest spatial frequencies, a compact interferometric imager and a conventional imaging instrument have a very similar behavior with respect to noise. For the former, the propagation of noise is unfavorable for the lower frequencies. However it is quite possible, for a particular task (detection for example), to design the transfer function of a compact interferometric imager to make it higher at the spatial frequencies relevant to the task.


\section{Conclusion}



In this paper, we performed an analysis of the compact interferometric imager concept. We recalled basic rules for the field of view and the resolution, and gave a sizing rule for the spectral channel width of such an instrument: the field of view is limited to the diffraction spot of an aperture, the resolution is given by the maximum baseline of the instrument, and the relative spectral width of a channel must be less than the ratio between the size of an aperture and the maximum baseline. 

We then modeled the propagation of the measurement noise in a  compact interferometric imager and we were able to compare this propagation to that of a focal plane imaging instrument after restoration by an optimal linear filter (Wiener). We performed this analysis within the framework of an identical photonic budget for the two types of instruments, in order to obtain a fair comparison.

This analysis allowed us to show that the noise propagation was, for a compact interferometric imager recording each spatial frequency once, very similar to that of a multi-aperture focal plane imager with maximum resolution (or minimum redundancy)~\cite{Cassaing-a-18}, \ie with a quasi-flat transfer function.

Much work remains to be done to fully demonstrate the feasibility of the concept from a technological point of view. On the algorithmic side, an interesting avenue to explore is the reconstruction of the observed scene from a hybrid instrument~\cite{Debary-a-22} consisting of a conventional small telescope recording a continuum of low spatial frequencies complemented by a compact interferometric imager recording a set of discrete high spatial frequencies.

\bigskip
\appendix

\section{Error covariance matrix for the Wiener restoration of an image}
\label{annexe-matcoverr}

In matrix form, the image model writes:
\begin{equation}
    \bfi' = \matH \bfo' + \bfn ,
\end{equation}
and the  object restored by an optimal\footnote{in the sense of a minimum mean square error, which also happens to be, under Gaussian hypotheses, the Maximum \emph{A Posteriori} estimator.} linear estimator  writes $\hat{\bfo}' = \matW \bfi'$, with:
\begin{equation}
    \matW = \left(\matH^T \matC_n^{-1} \matH + \matC_o^{-1} \right)^{-1} \matH^T \matC_n^{-1} \quad\text{in the so-called information form},
\end{equation}
or 
\begin{equation}
    \matW = \matC_o \matH^T \left(\matH \matC_o \matH^T + \matC_n\right)^{-1}
     \quad\text{in the so-called covariance form}.
\end{equation}
This estimator has a zero mean bias. The covariance of the error $\bfepsilon\triangleq \hat{\bfo}'-\bfo$ writes:
\begin{equation}
    \Gamma = \left(\matH^T \matC_n^{-1} \matH + \matC_o^{-1} \right)^{-1} \;.
    \label{eq-Gamma}
\end{equation}
Let  $\matF$ be the matrix performing the DFT, defined in Eq.~\ref{TFD}. Let $\matM^H$ be the conjugated and transposed matrix, also called Hermitian conjugate or adjoint matrix, of a matrix $\matM$. 
The inverse DFT, in order to ensure that $\matF^{-1}\matF=\matF\matF^{-1}=\mathrm{I}$, is normalized in the following way: $\matF^{-1} = (1/N_{\text{pix}}^2).\matF^H$.
The covariance matrix of the DFT of the error, referred to as $\Gamma_{\matF\bfepsilon}$, is given by:
\begin{equation}
    \Gamma_{\matF\bfepsilon} = \matF \Gamma \matF^H 
    = N_{\text{pix}}^2 \matF \Gamma \matF^{-1} ,
\end{equation}
thus
\begin{equation}
    \Gamma_{\matF\bfepsilon}^{-1} 
    = \frac{1}{N_{\text{pix}}^2}
    \matF \Gamma^{-1}  \matF^{-1}  \;.
\end{equation}
According to Equation~\ref{eq-Gamma}, we have:
\begin{align}
\Gamma_{\matF\bfepsilon}^{-1} 
    &= \frac{1}{N_{\text{pix}}^2}
    \matF \left(\matH^T \matC_n^{-1} \matH + \matC_o^{-1}\right)   \matF^{-1} \nonumber \\
    &= \frac{1}{N_{\text{pix}}^2 \sigma_n^2}\matF \matH^T \matH \matF^{-1}  
       + \frac{1}{N_{\text{pix}}^2} \matF \matC_o^{-1} \matF^{-1} \nonumber \\
    &= \frac{1}{N_{\text{pix}}^2 \sigma_n^2}\matF\matH^H\matF^{-1} \,\matF\matH \matF^{-1}  
       + \frac{1}{N_{\text{pix}}^2} \matF \matC_o^{-1} \matF^{-1} \quad\text{(because $\matH$ is real-valued)} \nonumber \\
    &= \frac{1}{N_{\text{pix}}^2 \sigma_n^2}(\matF\matH\matF^{-1})^H \,(\matF\matH \matF^{-1})  
       + \frac{1}{N_{\text{pix}}^2} (\matF \matC_o \matF^{-1})^{-1} \nonumber \\
    &= \frac{1}{N_{\text{pix}}^2 \sigma_n^2}(\matF\matH\matF^{-1})^H \,(\matF\matH \matF^{-1})  
       +  (\matF \matC_o \matF^H)^{-1}\;. \label{eq-Gamma-Fourier}
\end{align}
Because matrix $\matH$ expresses a convolution, it is  a Toeplitz-block-Toeplitz matrix, and thus approximately a circulant-block-circulant matrix. It is thus diagonalizable by a DFT with a good approximation, and its eigenvalues are values of the transfer function, $\bfhtilde$:
\begin{equation}
    \widetilde{\matH} \triangleq \matF\matH\matF^{-1} = \diag{\bfhtilde} \;.
    \label{eq-H-tilde}
\end{equation}
Besides, assuming that the object's prior probability distribution is second order stationary with Energy Spectral Density (ESD) $\bfS_o$, the prior covariance matrix of the DFT of the object is also diagonal, and this diagonal is equal to $\bfS_o$:
\begin{equation}
    \widetilde{\matS_o} \triangleq \matF \matC_o \matF^H = \diag{\bfS_o} \;.
    \label{eq-So-tilde}
\end{equation}
Using Equations~\ref{eq-H-tilde} and~\ref{eq-So-tilde} and injecting them into Equation~\ref{eq-Gamma-Fourier} yields:
\begin{align}
\Gamma_{\matF\bfepsilon}^{-1} 
    &= \frac{1}{N_{\text{pix}}^2 \sigma_n^2}\diag{|\bfhtilde|^2}  +\diag{\frac{1}{\bfS_o}}\nonumber\\
    &= \diag{\frac{1}{N_{\text{pix}}^2 \sigma_n^2} |\bfhtilde|^2 
    + \frac{1}{\bfS_o}} \nonumber\\
    &= \diag{\frac{1}{N_{\text{pix}}^2 \sigma_n^2} \left[|\bfhtilde|^2 
    + \frac{N_{\text{pix}}^2 \sigma_n^2}{\bfS_o} \right]} ,
 \label{eq-Gamma-Fourier2}
\end{align}
and finally:
\begin{equation}
\boxed{
    \Gamma_{\matF\bfepsilon} = 
    \diag{ \frac{N_{\text{pix}}^2 \sigma_n^2}{|\bfhtilde|^2 
    + \frac{N_{\text{pix}}^2 \sigma_n^2}{\bfS_o}} }
    } , \QED 
    \label{eq-Gamma-Fourier3}
\end{equation}
as announced in Equation~\ref{eq-mat-cov-err-F}.

\section*{ACKNOWLEDGEMENTS}  
This work has been partly funded by a \textsc{Cnes} contract (LMM, VM, FC). It has also been partly funded by Airbus Defence and Space (ADS) through HD's PhD fellowship.
We thank our colleagues Matthieu Castelnau, Mathieu Boutilier and Renaud Binet (\textsc{Cnes}) for fruitful discussions.

\clearpage
\bibliographystyle{spiebib}
\bibliography{Acronymes,EnglishAcronyms,Livres,Theses,Actes,Articles,BibLivres,laurent,Spider,spiderVSgalette,Spider_V}
\addcontentsline{toc}{section}{\refname}

\end{document}